\documentclass[sigconf,screen]{acmart}

\usepackage{CJKutf8}
\usepackage{cleveref}
\usepackage{algorithm}
\usepackage{algpseudocode}
\usepackage{adjustbox}
\usepackage{tikz}
\usetikzlibrary{calc, positioning, arrows.meta, shapes.geometric}
\AtBeginDocument{%
  }

\copyrightyear{2026}
\acmYear{2026}
\setcopyright{cc}
\setcctype{by}
\acmConference[ASE '26]{Proceedings of the 41st IEEE/ACM International Conference on Automated Software Engineering}{October 12--16, 2026}{Munich, Germany}
\acmBooktitle{Proceedings of the 41st IEEE/ACM International Conference on Automated Software Engineering (ASE '26), October 12--16, 2026, Munich, Germany}
\acmDOI{10.1145/3832783.3834407}
\acmISBN{979-8-4007-2882-2/2026/10}

\acmSubmissionID{ase26main-p2013-p}
\received{2026-03-26}
\received[accepted]{2026-06-18}

\begin{document}
\begin{CJK*}{UTF8}{min}

\title[Where Does Balance Break? Boundary Discovery for Game Balance Testing under a Finite Simulation Budget]{Where Does Balance Break? Boundary Discovery\\ for Game Balance Testing under a Finite Simulation Budget}

\author{Hiroki Mukai}
\email{is0692vs@ed.ritsumei.ac.jp}
\orcid{0009-0002-3939-0088}
\affiliation{%
  \institution{Ritsumeikan University}
  \city{Ibaraki}
  \state{Osaka}
  \country{Japan}
}

\author{Yusaku Kato}
\email{is0610ks@ed.ritsumei.ac.jp}
\orcid{0009-0007-5107-7111}
\affiliation{%
  \institution{Ritsumeikan University}
  \city{Ibaraki}
  \state{Osaka}
  \country{Japan}
}

\author{Norihiro Yoshida}
\email{norihiro@fc.ritsumei.ac.jp}
\orcid{0000-0003-4910-1729}
\affiliation{%
  \institution{Ritsumeikan University}
  \city{Ibaraki}
  \state{Osaka}
  \country{Japan}
}

\author{Erina Makihara}
\email{makihara@fc.ritsumei.ac.jp}
\orcid{0009-0008-8777-619X}
\affiliation{%
  \institution{Ritsumeikan University}
  \city{Ibaraki}
  \state{Osaka}
  \country{Japan}
}

\author{Katsuro Inoue}
\email{inoue-k@fc.ritsumei.ac.jp}
\orcid{0000-0001-5424-0614}
\affiliation{%
  \institution{Ritsumeikan University}
  \city{Ibaraki}
  \state{Osaka}
  \country{Japan}
}

\renewcommand{\shortauthors}{Mukai et al.}


\begin{abstract}
Software testing often relies on assumptions such as reproducible executions and stable correctness criteria.
However, many modern software systems exhibit non-deterministic executions and large behavior spaces, making exhaustive exploration impractical and single-run judgments unreliable.
These characteristics make it difficult to identify where acceptable behavior ends and problematic behavior begins.
Competitive multiplayer games represent a challenging instance of such systems, where balance must be maintained so that no single strategy dominates.
Even small parameter changes can trigger abrupt balance disruption, yet detecting such failures requires repeated simulations under non-deterministic outcomes and high-dimensional parameter spaces.
In this paper, we formulate game balance regression testing as a boundary-discovery problem under a finite simulation budget.
The objective is to efficiently identify inputs near the boundary that separates balanced and unbalanced regions.
To address this problem, we propose BBExplorer, which combines multi-directional candidate generation, budget-aware two-stage screening, and adaptive step-size shrinkage for boundary refinement.
Experimental results on two games with different levels of complexity show that the approach is strong in low-dimensional settings and remains effective in higher-dimensional ones.
It also exhibits stable boundary behavior across unseen random seeds and threshold settings.
These results indicate that BBExplorer is effective for practical balance regression testing and, more broadly, for boundary-oriented testing in non-deterministic, budget-constrained systems.
\end{abstract}





\maketitle

\section{Introduction}\label{sec:introduction}

Software testing has traditionally relied on assumptions such as reproducible executions and sufficiently stable correctness criteria, enabling systematic evaluation through coverage and test adequacy \cite{ammann2016introduction,Goodenough1975,Frankl1988,Bertolino2007}.
However, many modern software systems exhibit large behavior spaces and non-deterministic executions, making exhaustive exploration impractical and single-run judgments unreliable 
\cite{delemos2013software,Calinescu2018,Luo2014,Dutta2020,FSE2026Kato}.
In such systems, small variations in inputs or configurations can lead to disproportionately different behaviors, further complicating the identification of acceptable system behavior \cite{Fraser2013,Xiao2023}.

Competitive games represent a particularly challenging instance of such systems, where large behavior spaces and non-deterministic executions arise from complex player interactions.
Even small changes in parameters such as attack power or speed can disrupt balance, making certain strategies dominant and degrading player experience \cite{SweetserWyeth2005GameFlow,KangSuhKim2024MatchChurn}.
With over three billion players worldwide \cite{Newzoo2024,WhatsTheBigData2024}, such imbalances can affect a large user base.
This challenge is especially pronounced in live-ops games, where patches and updates continuously introduce new parameter configurations \cite{Dubois2021,murphy2014cowboys}.
For example, \textit{League of Legends} releases balance patches roughly every two weeks \cite{RiotGames2025Patch}, requiring developers to repeatedly assess the impact of parameter changes under time constraints.
To support regression testing during frequent game updates, GameRTS uses a state-transition-graph-based representation to select regression tests affected by game software changes \cite{GameRTS2023}.

These characteristics make balance evaluation inherently challenging under practical testing constraints.
In such settings, exhaustive evaluation of candidate inputs becomes unrealistic, while repeated simulations remain necessary because outcomes vary across runs.
As a result, developers need to understand not only which configurations perform well, but also where acceptable configurations stop and problematic ones begin.
From this perspective, game balance regression testing can be viewed as a boundary-discovery problem under non-deterministic evaluation and limited simulation budgets.

Existing research has addressed game balance from multiple directions, including restricted-play evaluation \cite{Jaffe2012}, player-behavior modeling \cite{Pfau2020, Pfau2024}, and simulation-driven optimization \cite{Rupp2025SimDriven, Budijono2022LUDUS, Hernandez2020}.
In practice, however, developers need more than a single recommended configuration.
They need to understand where balance disruption begins, namely, the boundary between balanced and unbalanced regions.
Boundary-near inputs provide this information.
They allow developers to verify whether a candidate patch remains on the safe side and to estimate how far parameters can be changed without introducing imbalance.
In practice, developers first determine an acceptable win-rate range from
their design goals and use its upper or lower limit as the target depending
on whether they are testing overpowered or underpowered behavior.
The discovered boundary-near inputs can then be retained as a patch-level
regression suite and rerun after updates; inputs that move into the unbalanced
region become candidates for inspection, rebalancing, or CI guardrail tests.
In contrast to an optimum, which provides a single solution, boundary information characterizes the extent and shape of the safe region, including directions in which updates become risky.
Identifying such boundaries is challenging.
The parameter space is multi-dimensional, and boundaries may extend in multiple directions, making them difficult to capture with a single search trajectory.
Moreover, game testing is already time-consuming in practice \cite{murphy2014cowboys, Politowski2021, Santos2018}, and each balance evaluation requires repeated simulations to obtain reliable win-rate estimates.
As a result, exhaustive evaluation of candidate configurations is impractical within a single patch cycle.

From a software testing perspective, this boundary-discovery task is better viewed as a test-input selection and generation problem \cite{Bertolino2007, McMinn2004SBSTSurvey, AnandEtAl2013TestGenSurvey} within a finite simulation budget, rather than an optimization problem that seeks a single best configuration.
The input space is high-dimensional, win-rate measurement is non-deterministic \cite{Luo2014}, and each candidate input requires repeated executions.
As a result, dense evaluation of all configurations is infeasible \cite{Fraser2013, Politowski2021}, making efficient budget allocation across candidate inputs a central concern \cite{JamiesonTalwalkar2016SH, LiEtAl2018Hyperband}.


In this paper, we formulate game balance regression testing \cite{YooHarman2012RegressionSurvey} as a boundary-discovery problem under non-deterministic evaluation and a finite simulation budget.
The objective is to efficiently explore the neighborhood of the boundary that separates balanced and unbalanced regions.
Unlike prior optimization-based approaches, which focus on identifying a single optimal configuration, our goal is to explicitly characterize the boundary at which acceptable behavior breaks down.
To address this objective, we introduce \textbf{BBExplorer} (Balance Boundary Explorer), which combines multi-directional candidate generation, two-stage candidate screening, and adaptive step-size shrinkage for boundary refinement.


The contributions of this paper are as follows:
\begin{enumerate}
    \item We formulate game balance regression testing as a boundary-discovery
problem for generating test inputs under non-deterministic evaluation
and a finite simulation budget.
    \item We propose \textbf{BBExplorer}, which integrates multi-directional candidate generation, two-stage candidate screening, and adaptive step-size shrinkage for boundary refinement.
    \item Through experiments on a custom turn-based game (\textit{Turnbased}) and a real-time strategy game (\textit{Generals}), we demonstrate that \textbf{BBExplorer} achieves effective boundary localization in both low- and higher-dimensional settings and remains robust under seed and threshold variations.
\end{enumerate}

\section{Motivating Example}
\label{sec:motivating-example}

\subsection{Win-Rate Terrain}
To motivate our approach, we use a simple turn-based competitive game with a two-dimensional parameter space as a running example.
In this game, a focal character competes against a fixed opponent in a turn-based duel.
The focal character has two adjustable parameters: attack power (\textbf{ATK}), which influences damage, and speed (\textbf{SPD}), which affects turn order and evasion.

\begin{figure}[t]
  \centering
  \includegraphics[width=0.78\linewidth]{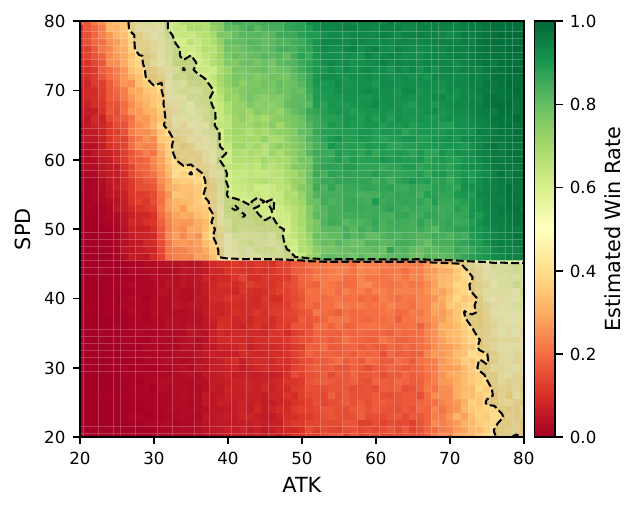}
  \caption{Estimated win-rate terrain over ATK and SPD.
  Colors indicate estimated win rates; the dashed band marks
  the acceptable range (0.4--0.6) separating balanced and unbalanced regions.}
  \Description{Heatmap of estimated win rate over ATK and SPD, generated by sweeping the focal character's parameters against a fixed opponent through repeated simulations. A curved boundary band separates balanced and unbalanced regions, with steep color transitions indicating cliffs.}
  \label{fig:winrate-terrain}
\end{figure}

\Cref{fig:winrate-terrain} visualizes how the focal character's win rate changes over the ATK--SPD space.
We refer to this heatmap of estimated win rates as the win-rate terrain.
To obtain it, we fix the opponent-side evaluation condition and sweep the focal character's ATK and SPD over a two-dimensional grid.
At each grid point, repeated simulations are executed to obtain an estimated win rate, and the resulting values are visualized over the ATK--SPD grid.
From this figure, several intuitive characteristics of boundary-near behavior can be observed.
First, between balanced and unbalanced regions, we observe a transitional boundary band where the win rate can easily cross the acceptable threshold.
Second, within parts of this band, small parameter changes produce abrupt shifts in win rate.
In this subsection, we refer to such locally steep regions as \textbf{cliffs} in an observation-based sense.
Third, the transitional region is curved rather than linear, suggesting that ATK and SPD interact in shaping the boundary.
We use these terms as descriptive labels for the visible phenomena here, and formalize them later.

These observations suggest that balance disruption begins along a curved and locally fragile boundary-near region rather than at an isolated point.
They also suggest why discovering such regions is difficult: simple single-axis exploration or coarse evaluation may fail to capture where the balance judgment actually changes.
The next subsection explains this difficulty from the perspectives of interaction, non-determinism, and budget constraint.
\label{sec:winrate-terrain}
\subsection{Why Boundary Discovery Is Hard}
\label{sec:why-boundary-hard}

The terrain in \Cref{fig:winrate-terrain}, introduced in \Cref{sec:winrate-terrain}, suggests why boundary discovery is difficult.
This difficulty can be understood from three aspects: interaction, non-determinism, and budget constraints.

First, interactions make it difficult to infer the boundary shape from a single axis.
When multiple parameters (e.g., ATK and SPD) jointly affect the win rate, the boundary can take on a curved shape.
As a result, single-axis exploration is insufficient and can miss regions near the boundary.

Second, win-rate measurement is non-deterministic. Even with the same parameter configuration, match results vary due to randomness and opponent behavior. Reliable win-rate estimation, therefore, requires repeated executions rather than a single run. This instability becomes especially important near the boundary, where small estimation errors can easily flip the judgment between balanced and unbalanced.

Third, dense evaluation used to produce \Cref{fig:winrate-terrain} is realistic only in a low-dimensional example. Because each point requires repeated executions for reliable win-rate estimation, a dense grid can become prohibitively expensive as dimensionality increases. Under a finite simulation budget, the central question is therefore how to allocate evaluation effort across candidate inputs, rather than how to evaluate the whole space exhaustively.

These difficulties directly translate into the design requirements introduced next.

\subsection{Requirements}\label{sec:requirements}

In this subsection, we formulate three design requirements (R1, R2, and R3) for a practical boundary discovery method.
Each requirement addresses one of the difficulties discussed above; because these difficulties are closely connected, the three requirements must ultimately be considered together.

\paragraph{R1: Multi-directional exploration.}
Due to interactions, the boundary may bend in complex ways in a multi-dimensional space. As a result, exploration along only a single axis may capture the boundary only partially.
Both its position and its shape may therefore be characterized incompletely. A boundary discovery method is therefore required to generate candidate points in multiple directions and explore the space from multiple perspectives.

\paragraph{R2: Sampling for judgment reliability.}
Near the boundary, an estimated win rate obtained from only a small number of executions can easily move across the threshold, making pass/fail judgment unstable.
What matters here is not simply increasing the number of evaluations, but ensuring sufficient samples at each candidate point to support a reliable judgment.
In other words, R2 is needed to make boundary detection reliable.

\paragraph{R3: Efficient budget allocation.}
In practice, balance testing already consumes substantial development effort~\cite{murphy2014cowboys, Politowski2021}, and the per-candidate cost of repeated match evaluation further tightens this constraint.
At the same time, it is unrealistic to apply such a reliable evaluation uniformly to all candidate points under a limited computational budget.
The method must therefore allocate more budget to points that appear promising and likely near the boundary, while suppressing excessive evaluation of points that are likely far from it.
In other words, R3 is not about improving judgment reliability itself, but about deciding where limited resources should be spent.

These three requirements must be satisfied simultaneously.
Multi-directional exploration (R1) increases the number of candidate points, while sampling for judgment reliability (R2) increases the evaluation cost per point.
Trying to satisfy both naively would quickly exhaust the computational budget.
Because R1, R2, and R3 must be satisfied simultaneously under a limited budget, budget allocation (R3) becomes the central design problem: it determines which candidates should be screened coarsely and which should be examined in detail. 

To address this trade-off, our proposed method, \textbf{BBExplorer},
combines multi-directional candidate generation, budget-aware
two-stage screening, and step-size shrinkage, as detailed in
\Cref{sec:approach}.

\section{Problem Formulation}
\label{sec:problem}

\subsection{Parameter Space and Win Rate}
\label{sec:param-winrate}

In this section, we formally define the boundary-discovery problem in game balance.

Let $p$ denote the complete set of parameters that determine the outcome of a competitive game match.
We decompose $p$ into \textbf{controllable parameters} $P_c$ and \textbf{uncontrollable parameters} $P_{\bar{c}}$, and write $p = (P_c, P_{\bar{c}})$.
Here, $P_c \subseteq \mathbb{R}^d$ is the controllable parameter space, where each dimension corresponds to a game parameter (e.g., attack power ATK, speed SPD, or defense \textbf{DEF}).
A point in this space, $P_c^{(i)} = (p_{c,1}^{(i)}, \ldots, p_{c,d}^{(i)})$, represents one concrete parameter configuration.
By contrast, $P_{\bar{c}} \subseteq \mathbb{R}^m$ denotes environment-side factors that affect match outcomes but are not themselves search variables, such as the opponent's parameter configuration, map layout, and random seed governing stochastic match outcomes.

For brevity, in the remainder, the simplified notation $P_c = P$, $P_{\bar{c}} = \bar{P}$, $P_c^{(i)} = P^{(i)}$, and $p_{c,k} = p_k$ is used.
Under a fixed $\bar{P}$, the induced \textbf{win-rate function} over the controllable space is written as
\begin{equation}
    f_p(\cdot \mid \bar{P}) : P \to [0,1]
\end{equation}
where $\cdot$ is a placeholder for the controllable configuration to be evaluated.
Accordingly, the win rate at a configuration $P^{(i)}$ is written as $f_p(P^{(i)} \mid \bar{P})$.
This notation makes explicit that environment-side factors are treated as a fixed evaluation condition $\bar{P}$, while $P^{(i)}$ is the object explored by a developer.

This interpretation also matches the operational meaning used in the experiments.
Within one search run, $\bar{P}$ should be read as a fixed evaluation condition in the simulator: the opponent-side setup, the simulation condition, and the distribution of stochastic outcomes are treated as fixed, while \textbf{BBExplorer} updates only $P$.
Accordingly, the search changes only the controllable parameters, and the environment-side condition remains the fixed evaluation condition.

In practice, the true value of $f_p(P^{(i)} \mid \bar{P})$ is generally unavailable analytically.
For a given parameter configuration $P^{(i)}$, we therefore run $N$ simulations and define the \textbf{estimated win rate} $\hat{f}_p(P^{(i)} \mid \bar{P})$ as
\begin{equation}
    \label{eq:estimated-winrate}
    \hat{f}_p(P^{(i)} \mid \bar{P}) = \frac{1}{N} \sum_{j=1}^{N} \mathbb{I}[\text{win in match } j]
\end{equation}
Here, $\mathbb{I}[\cdot]$ is the indicator function, which returns 1 when the condition holds and 0 otherwise.
This is an empirical estimate of the win rate under the fixed condition $\bar{P}$.
Under standard repeated-sampling assumptions, $\hat{f}_p(P^{(i)} \mid \bar{P})$ becomes a more stable approximation of $f_p(P^{(i)} \mid \bar{P})$ as $N$ increases.
When $N$ is small, however, the estimate can vary substantially, making the resulting balance judgment unstable, especially near a decision threshold.

The robustness analyses in \Cref{sec:robustness} follow the same distinction.
A seed shift tests what happens when the stochastic aspect of the fixed condition $\bar{P}$ changes.
A threshold shift, by contrast, changes the boundary definition.
The evaluation budget is a separate testing-side setting: it determines how thoroughly the search can be conducted but does not alter the boundary or the environment condition.
Throughout this paper, we therefore distinguish environment-side conditions represented by $\bar{P}$ from testing-side settings such as the threshold and the evaluation budget.

\subsection{Balanced Set and Boundary}
\label{sec:balanced-set}

In this subsection, we define the balanced set and its boundary under a fixed $\bar{P}$.
In the remainder, $\bar{P}$ denotes the fixed evaluation condition for one search run, whereas the threshold and evaluation budget are testing-side settings.

Let $[l,h] \subseteq [0,1]$ be the acceptable win-rate interval.
Under a given $\bar{P}$, we define the \textbf{balanced set} as
\begin{equation}
\mathcal{B}_{\bar{P}} = \{ P^{(i)} \in P : l \le f_p(P^{(i)} \mid \bar{P}) \le h \}
\end{equation}
This is the set of parameter configurations regarded as balanced under that evaluation condition.

The corresponding boundary is given by $\partial \mathcal{B}_{\bar{P}}$, where $\partial$ denotes the boundary operator.
This boundary marks where the judgment switches between balanced and unbalanced configurations.
The \textbf{boundary band} introduced in \Cref{sec:winrate-terrain} should be understood not as the formal boundary itself, but as an informal transition region around the formal boundary.

In practice, the goal of this paper is not exhaustive characterization of all balanced configurations, but efficient discovery of boundary-near inputs under a finite simulation budget.
Thus, while the balanced set and boundary are defined under a fixed $\bar{P}$, the practical interest lies in capturing the neighborhood of that boundary efficiently.

The next two subsections refine this view: \Cref{sec:sensitivity-cliff} introduces sensitivity and cliffs near the boundary, and \Cref{sec:testing-goal} formulates the resulting testing goal under budget.

\subsection{Sensitivity and Cliff}
\label{sec:sensitivity-cliff}

In this subsection, we introduce the notions of sensitivity and cliff to describe steep local changes near the boundary.
The boundary marks where the judgment switches between balanced and unbalanced configurations.
A cliff, in contrast, denotes a fragile region where the win rate changes sharply under small parameter perturbations.
These two notions are often close in practice, but they play different conceptual roles.

Developers directly observe and modify only the controllable space $P$.
Under a fixed $\bar{P}$, the visible win-rate terrain can therefore be treated as one slice of the full space.
To capture local change on this slice, we define the sensitivity under a given $\bar{P}$ as
\begin{equation}
S(P^{(i)} \mid \bar{P}) = \| \nabla_P f_p(P^{(i)} \mid \bar{P}) \|
\end{equation}
where $\nabla_P$ denotes the gradient with respect to the controllable parameters.
This quantity measures how strongly the expected win rate responds to small parameter changes.

Next, let $\tau > 0$ be a cliff threshold.
We define the cliff set as
\begin{equation}
\mathcal{C}_{\bar{P}} = \{ P^{(i)} \in P : S(P^{(i)} \mid \bar{P}) \ge \tau \}
\end{equation}
This set represents regions where small parameter changes can induce large win rate shifts.
In this paper, the cliff concept is used primarily to explain why progressively finer boundary localization is needed near steep local transitions, rather than as a directly optimized or separately evaluated target.

\begin{figure}[t]
  \centering
  \includegraphics[width=0.8\linewidth]{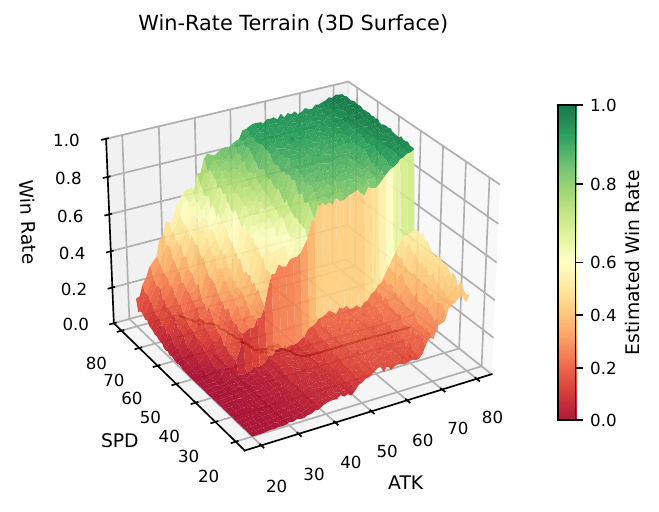}
  \caption{Three-dimensional view of the win-rate terrain. Steep surface regions correspond to cliff-like areas with large local sensitivity.}
  \Description{Three-dimensional surface plot of the win-rate terrain. Steep slopes indicate regions where small parameter changes cause large win-rate shifts.}
  \label{fig:cliff-visualization}
\end{figure}

\Cref{fig:cliff-visualization} provides an intuitive view of a cliff.
The steep regions suggested by the sharp color transitions in \Cref{fig:winrate-terrain} correspond to the formal cliff concept introduced here.
Later experiments revisit these steep regions through the boundary-near inputs discovered by \textbf{BBExplorer}.

\subsection{Testing Goal under Budget}
\label{sec:testing-goal}

Building on the preceding definitions, we now clarify the testing goal addressed in this paper.
Unlike conventional software testing, where expected outputs can often be specified before execution, in game balance testing the expected win rate for a test input $P^{(i)}$ is unavailable a priori.
The estimated win rate under a fixed $\bar{P}$, written as $\hat{f}_p(P^{(i)} \mid \bar{P})$, can only be obtained after actually executing simulations.
This \textbf{execution-based judgment} forms the basis for the boundary-discovery problem addressed in this paper.

We first introduce the \textbf{balance judgment condition} only to fix the decision criterion.
For a test input $P^{(i)} \in P$,
\begin{equation}
    l \le \hat{f}_p(P^{(i)} \mid \bar{P}) \le h
\end{equation}
means that the input is judged balanced.

The main goal of this paper is not to confirm whether an input is balanced.
We seek to efficiently obtain boundary-near inputs within a finite simulation budget.
Recovering the full boundary via dense-grid evaluation requires many simulations, even in low-dimensions, and becomes even less realistic in higher dimensions.
Under a finite simulation budget, a practical objective is therefore to generate as many boundary-near inputs as possible.

From this perspective, this paper formulates boundary discovery as a \textbf{test input generation} problem.
Given a target boundary value $b \in \{l,h\}$ and a tolerance $\epsilon > 0$, a test input $P^{(i)}$ is regarded as boundary-near if it satisfies
\begin{equation}
    \label{eq:boundary-near-condition}
    |\hat{f}_p(P^{(i)} \mid \bar{P}) - b| \le \epsilon
\end{equation}

\paragraph{boundary-discovery problem.}
Given a finite computational budget $B$ and a starting point set $X_0 = \{x_0^{(1)}, \ldots, x_0^{(K)}\}$, starting from each point in $X_0$, generate as many test inputs $P^{(i)}$ satisfying the boundary-near condition in \Cref{eq:boundary-near-condition} as possible.

\textbf{BBExplorer} aims to generate boundary-near test inputs automatically.
The resulting set of inputs provides developers with practical information about the boundary's position and shape, supporting balance tuning within a limited budget.

\section{Approach}
\label{sec:approach}

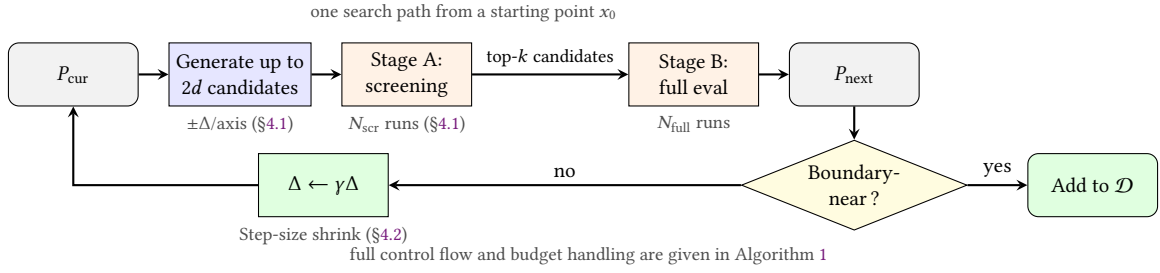
\begin{figure*}[!t]
  \centering
  \begin{tikzpicture}[
    >=stealth,
    scale=0.88, transform shape,
    state/.style={
      draw, rounded corners=4pt,
      minimum height=0.95cm, minimum width=1.95cm,
      align=center, font=\normalsize, fill=gray!10
    },
    store/.style={
      draw, rounded corners=4pt,
      minimum height=0.95cm, minimum width=1.95cm,
      align=center, font=\normalsize, fill=green!12
    },
    proc/.style={
      draw,
      minimum height=0.95cm, minimum width=1.95cm,
      align=center, font=\normalsize
    },
    dec/.style={
      draw, diamond, aspect=2.5,
      align=center, font=\normalsize,
      fill=yellow!15, inner sep=2pt
    },
    note/.style={font=\small, text=gray!60!black, align=center}
  ]

    \node[state] (pcur) {$P_{\mathrm{cur}}$};

    \node[proc, fill=blue!10, right=0.45cm of pcur] (gen)
      {Generate up to\\$2d$ candidates};
    \node[note, below=1pt of gen]
      {$\pm\Delta$/axis (\S\ref{sec:multi-directional})};

    \node[proc, fill=orange!10, right=0.45cm of gen] (stA)
      {Stage A:\\screening};
    \node[note, below=1pt of stA]
      {$N_{\text{scr}}$ runs (\S\ref{sec:two-stage-eval})};

    \node[proc, fill=orange!10, right=2.35cm of stA] (stB)
      {Stage B:\\full eval};
    \node[note, below=1pt of stB]
      {$N_{\text{full}}$ runs};

    \node[state, right=0.45cm of stB] (pnext) {$P_{\mathrm{next}}$};

    \draw[->, thick] (pcur) -- (gen);
    \draw[->, thick] (gen) -- (stA);
    \draw[->, thick] (stA) -- node[above, font=\small] {top-$k$ candidates} (stB);
    \draw[->, thick] (stB) -- (pnext);

    \node[note, above=6pt of $(pcur.north)!0.5!(pnext.north)$]
      {one search path from a starting point $x_0$};

    \node[dec, below=0.5cm of pnext] (check) {Boundary-\\near\,?};
    \draw[->, thick] (pnext) -- (check);

    \node[store, right=0.9cm of check] (addD) {Add to $\mathcal{D}$};
    \draw[->, thick] (check) -- node[above, font=\normalsize] {yes} (addD);

    \coordinate (shrinkX) at ($(gen)!0.5!(stA)$);
    \node[proc, fill=green!12] (shrink) at (shrinkX |- check)
      {$\Delta \gets \gamma\Delta$};
    \node[note, below=1pt of shrink]
      {Step-size shrink (\S\ref{sec:step-shrinkage})};

    \draw[->, thick] (check.west) -- node[above, font=\normalsize] {no} (shrink.east);
    \draw[->, thick] (shrink.west) -| (pcur.south);

    \node[note] at ($(shrink.south)!0.5!(check.south)+(0,-0.48)$)
      {full control flow and budget handling are given in \Cref{alg:adaptive-boundary-search}};

  \end{tikzpicture}
  \caption{Overview of \textbf{BBExplorer}. The top row shows one iteration of a search path: candidate generation, screening, and selection of the next center. The bottom row shows the update loop: if the selected center is boundary-near, it is added to $\mathcal{D}$; otherwise, the step size is shrunk and the search continues.}
  \Description{Two-row workflow diagram of \textbf{BBExplorer}.
    Top row: current search center, multi-directional candidate generation,
    Stage A screening, top-k selection, Stage B full evaluation, and next
    search center.
    Bottom row: boundary-nearness check; if yes, the point is added to the
    result set; if no, the step size is shrunk and the iteration loops back.}
  \label{fig:method-overview}
\end{figure*}

\textbf{BBExplorer} (Balance Boundary Explorer), introduced in
\Cref{sec:introduction}, is an iterative approach for the boundary-discovery
problem defined in \Cref{sec:problem} under a finite simulation budget.
To address R1--R3 (\Cref{sec:requirements}), it combines
(i) multi-directional candidate generation,
(ii) budget-aware two-stage screening, and
(iii) step-size shrinkage.
These components play complementary roles:
multi-directional candidate generation probes positive and negative directions
along each dimension (R1); two-stage screening reduces stochastic uncertainty
through repeated simulations (R2); and top-$k$ full evaluation concentrates
the finite simulation budget on candidates estimated to be closest to the
target boundary (R3).
\Cref{fig:method-overview} and \Cref{alg:adaptive-boundary-search}
summarize the overall procedure.

From each $x_0 \in X_0$, \textbf{BBExplorer} uses an equal budget share $B/|X_0|$.
At each iteration, it generates candidates (Line~5), selects the next center by two-stage screening (Line~6), and checks boundary-nearness.
If the selected center is boundary-near, it is added to $\mathcal{D}$ and that path terminates (Lines~8--10); otherwise, the step size is shrunk and the search continues (Line~13).

We describe the rationale and details of each component.

\subsection{Candidate Generation and Screening}
\label{sec:multi-directional}
\label{sec:two-stage-eval}

This subsection describes one iteration from candidate generation to next-center selection (\Cref{alg:adaptive-boundary-search}, Lines~5--6).

\paragraph{Multi-directional candidate generation.}
Let $x_t \in P$ be the current search center at iteration $t$ and let $\Delta_t$ be the current step size.
For each basis vector $e_i$, it considers $x_t + \Delta_t e_i$ and $x_t - \Delta_t e_i$, and clips the result to the feasible range of $P$ if it falls outside.
The number of candidates per iteration is therefore at most $2d$.
This design directly addresses R1 (multi-directional exploration) in \Cref{sec:requirements}.
Probing both positive and negative directions along every axis improves robustness to local boundary-shape changes, while keeping the number of candidates linear in dimension.
It also enables direct comparison of directions by boundary proximity before choosing the next center.

\paragraph{Budget-aware two-stage screening.}
Given the candidate set $\mathcal{C}$, \textbf{BBExplorer} allocates simulation budget in two stages under the same fixed condition $\bar{P}$ (\Cref{fig:method-overview}).
As discussed in \Cref{sec:param-winrate}, reliable boundary detection under non-determinism requires a sufficient number of simulation runs per candidate.
However, allocating that cost to every candidate is infeasible under a limited budget.
\textbf{BBExplorer} addresses this trade-off in two stages.

\textbf{Stage~A: Screening.}
The role of this stage is coarse screening by estimated boundary proximity, not the final judgment.
For all $2d$ candidates, \textbf{BBExplorer} executes a small number of simulations, $N_{\text{scr}}$, and obtains a preliminary estimated win rate $\hat{f}_p^{\text{scr}}(\mathbf{c} \mid \bar{P})$.
The screening score of a candidate $\mathbf{c}$ is defined as
\begin{equation}
    \label{eq:screening-score}
    s(\mathbf{c}) = \frac{1}{\left|\hat{f}_p^{\text{scr}}(\mathbf{c} \mid \bar{P}) - b\right| + \varepsilon_s}
\end{equation}
where $b$ is the target boundary value and $\varepsilon_s > 0$ is a small constant to prevent division by zero.

Based on these scores, \textbf{BBExplorer} selects the top-$k$ candidates:
\begin{equation}
    \label{eq:screen-selection}
    \mathcal{C}^{\text{scr}} = \operatorname{Screen}(\mathcal{C}, s, k)
\end{equation}
Only the top-$k$ candidates advance to Stage~B, concentrating the remaining budget on the most promising boundary-near directions.

\textbf{Stage~B: Full evaluation.}
The role of this stage is to determine the next search center based on reliable win-rate estimates.
For the selected $k$ candidates, \textbf{BBExplorer} executes $N_{\text{full}}$ simulations ($N_{\text{full}} \gg N_{\text{scr}}$) and computes $\hat{f}_p^{\text{full}}(\mathbf{c} \mid \bar{P})$.
The candidate closest to the target boundary value is selected as the next search center.
Because this decision rests on $N_{\mathrm{full}}$ runs rather than on
the coarse screening estimate, the center update uses a more precise
win-rate estimate than the screening stage.

This two-stage design addresses R2 (sampling for judgment reliability) and R3 (efficient budget allocation) simultaneously.
It allocates more simulations to promising candidates while avoiding
expensive full evaluation of clearly non-promising directions.

The computational advantage is also clear.
While full evaluation of all candidates costs $2d \times N_{\text{full}}$ simulations, the two-stage procedure costs
\begin{equation}
    2d \times N_{\text{scr}} + k \times N_{\text{full}}
\end{equation}
which is substantially smaller when $k \ll 2d$ and $N_{\text{scr}} \ll N_{\text{full}}$.

\begin{algorithm}[!t]
\caption{\textbf{BBExplorer}}
\label{alg:adaptive-boundary-search}
\begin{algorithmic}[1]
\Require Starting point set $X_0$, fixed evaluation condition $\bar{P}$, target boundary value $b$, tolerance $\epsilon$, initial step size $\Delta_0$, decay factor $\gamma$, budget $B$, screening size $k$
\Ensure Set of discovered boundary-near inputs $\mathcal{D}$
\State $\mathcal{D} \gets \emptyset$
\For{each starting point $x_0 \in X_0$}
    \State $P_{\mathrm{cur}} \gets x_0$, $\Delta \gets \Delta_0$, $b_{\mathrm{used}} \gets 0$
    \While{$b_{\mathrm{used}} < B / |X_0|$}
        \State $\mathcal{C} \gets \textsc{GenerateCandidates}(P_{\mathrm{cur}}, \Delta)$ \Comment{\Cref{sec:multi-directional}: $2d$ candidates}
        \State $P_{\mathrm{next}} \gets \textsc{TwoStageScreen}(\mathcal{C}, \bar{P}, b, k)$ \Comment{\Cref{sec:two-stage-eval}}
        \State Update $b_{\mathrm{used}}$ with simulation cost
        \If{$|\hat{f}_p(P_{\mathrm{next}} \mid \bar{P}) - b| \le \epsilon$}
            \State $\mathcal{D} \gets \mathcal{D} \cup \{P_{\mathrm{next}}\}$
            \State \textbf{break} \Comment{move to next starting point}
        \EndIf
        \State $P_{\mathrm{cur}} \gets P_{\mathrm{next}}$
        \State $\Delta \gets \gamma \cdot \Delta$ \Comment{\Cref{sec:step-shrinkage}}
    \EndWhile
\EndFor
\State \Return $\mathcal{D}$
\end{algorithmic}
\end{algorithm}

\subsection{Step-Size Shrinkage}
\label{sec:step-shrinkage}

In this subsection, we describe the update rule for shrinking the step size $\Delta$ as the search progresses (\Cref{alg:adaptive-boundary-search}, Line~13).
The role of this step is to combine coarse exploration and fine localization within a single search process.

A fixed step size involves a fundamental trade-off.
A large step size may overshoot boundary-near fragile regions, especially cliff-like regions (with sharp win-rate drops), leading to coarse boundary approximation.
Conversely, a small step size requires many iterations to reach the boundary from distant starting points, wasting the budget.
The desirable step size therefore differs between early coarse exploration and later fine localization.

To resolve this trade-off, \textbf{BBExplorer} monotonically decreases the step size as iterations progress.
Specifically, it applies exponential decay with a decay factor $\gamma \in (0,1)$, so that the step size is updated as $\Delta_{t+1} = \gamma \Delta_t$,

where $\Delta_t$ is the step size at iteration $t$.
This update allows \textbf{BBExplorer} to approach the boundary quickly in early iterations and then localize boundary-near regions more precisely in later iterations.

Step-size shrinkage is more than decay.
Its main effect is to avoid both overshooting and stagnation while enabling a transition from coarse exploration to fine-grained refinement.
This transition is especially important near the cliff-like steep regions discussed in \Cref{sec:sensitivity-cliff}, where a coarse step can jump over relevant boundary-near regions.

To prevent the step size from becoming excessively small, a lower bound $\Delta_{\min}$ can be introduced, and the step size can be clipped when $\Delta_t < \Delta_{\min}$.
However, the essential point is the coarse-to-fine change in search resolution rather than the lower bound itself.
The specific hyperparameter values are reported in \Cref{sec:experiments}.

\section{Experiments}
\label{sec:experiments}

This section presents an experimental study of \textbf{BBExplorer} under finite simulation budgets from four perspectives: low-dimensional performance, higher-dimensional performance, robustness under seed and threshold shifts, and computational cost.

Our experiments compare \textbf{BBExplorer} with \textit{Random} and \textit{Fixed-Delta}.
\textit{Random} is a uniform-sampling baseline, whereas \textit{Fixed-Delta} is an ablation that removes step-size shrinkage from \textbf{BBExplorer}.
This comparison allows us to separate the effect of structured search itself from the additional benefit of coarse-to-fine refinement.
We refer to \textit{Random} and \textit{Fixed-Delta} collectively as the baselines throughout this section.

We address the following research questions.
\begin{itemize}
    \item RQ1: Is \textbf{BBExplorer} more effective than the baselines in terms of discovery rate and boundary proximity?
    \item RQ2: To what extent does \textbf{BBExplorer} retain its discovery performance as dimensionality increases?
    \item RQ3: Do the boundary-near inputs discovered by \textbf{BBExplorer} remain stable under different random seeds and boundary thresholds?
\end{itemize}


\subsection{Experimental Setup}
\label{sec:exp-setup}
\label{sec:environments}
\label{sec:exp-environments}
\label{sec:metrics}
\label{sec:exp-metrics}


We use two contrasting environments in the experiments: \textit{Turnbased}, a controlled low-dimensional environment, and \textit{Generals}, a demanding high-dimensional environment. Together with the corresponding simulator settings and evaluation metrics, they allow us to evaluate both the basic behavior of \textbf{BBExplorer} in controlled low-dimensional settings and its robustness in more demanding ones.

\textit{Turnbased} is based on a custom turn-based competitive game in which the boundary shape is relatively simple.
It is mainly used in 2D and 3D conditions to assess the basic performance of \textbf{BBExplorer} in low-dimensional settings.

\textit{Generals} is based on the real-time strategy game of the same name\footnote{\url{https://generals.io}}~\cite{straka2025artificial}, implemented on top of an adapted version of the open-source simulator\footnote{\url{https://github.com/strakam/generals-bots}}.
This environment is higher-dimensional, more computationally expensive, and subject to stronger parameter interactions, making it a more demanding test bed for \textbf{BBExplorer}.
In this paper, we use it mainly in 3D, 4D, and 5D conditions to study higher-dimensional behavior.

Across both environments, all methods are evaluated under the same environment-specific simulation budget and the same match-evaluation protocol.
Each search run is conducted under a fixed $\bar{P}$; this setting serves as the reference point for the seed-shift and threshold-shift analyses in \Cref{sec:exp-results}.

We do not claim that the following settings are optimal; they instantiate
the testing target, cost--reliability trade-off, and search resolution used
in our evaluation.\footnote{
Boundary setting:
$\epsilon=0.01$,
$b=0.60$ in the main experiments,
$b\in\{0.40,0.50,0.60\}$ in the threshold-shift analysis,
and $k=1$.
\textit{Turnbased} 3D:
$N_{\mathrm{scr}}=200$,
$N_{\mathrm{full}}=2000$,
$\Delta_0=15.0$,
$\gamma=0.90$,
$\Delta_{\min}=1.0$,
and $\texttt{max\_iter}=50$.
\textit{Generals}:
$N_{\mathrm{scr}}=20$,
$N_{\mathrm{full}}=200$,
$\gamma=0.92$,
and $\texttt{max\_iter}=40$;
$\Delta_0/\Delta_{\min}
=[6.0,1.6,0.7]/[0.5,0.1,0.05]$ in 3D,
with $2.0/0.15$ and $1.2/0.05$ appended in 4D and 5D, respectively.
Starts and trials:
fixed-seed C-LHS with seed 42 for \textit{Turnbased}
and seed 0 for \textit{Generals};
30 trials for \textit{Turnbased} 2D/3D and \textit{Generals} 5D,
20 for \textit{Generals} 3D/4D,
and 10 for the additional $b=0.40$ and $b=0.50$ settings.
Implementation:
\texttt{max\_iter} capped each search path,
\texttt{total\_games} recorded the consumed matches corresponding to $B$,
and each candidate evaluation used
$N_{\mathrm{scr}}$ or $N_{\mathrm{full}}$ repeated matches.
}

We use the following evaluation metrics, organized into four roles: success, efficiency, boundary proximity, and search-process quality. Let $|D|$ denote the number of discovered boundary-near inputs returned from the $|X_0|$ search paths, and let $M$ denote the total number of matches consumed.
As the success metric, we use \textbf{discovery rate} $R = |D| / |X_0|$: the fraction of starting points from which the method reaches a point satisfying the boundary-near condition in \Cref{eq:boundary-near-condition}.
As the efficiency metric, we use \textbf{discovery efficiency} $E = 10^4 |D| / M$: the number of boundary discoveries per 10{,}000 matches consumed.
In sparse-hit settings, $E$ is not interpreted alone but used alongside $R$ and the proximity metrics.

As the boundary-proximity metric, we use \textbf{mean boundary distance} $\bar{D}$: the mean absolute difference between the estimated win rate at the final point of each search path and the target boundary value.
It measures how close the method gets to the boundary by the end of the search, even when it does not converge.

As the search-process-quality metric, we use \textbf{boundary distance AUC} $A$: the area under the per-iteration boundary-distance curve over the search process.
Unlike $\bar{D}$, which captures only the endpoint proximity, $A$ reflects whether the search trajectory stays close to the boundary from early stages onward.

For statistical comparison, we use the Mann--Whitney U test and report Vargha--Delaney's $\hat{A}_{12}$ as the effect size.
This combination allows us to evaluate not only whether a difference exists, but also how large it is.


\subsection{Baselines}
\label{sec:exp-baselines}

\textit{Random} samples candidate points uniformly without directional guidance, serving as an unstructured baseline.
In contrast, \textit{Fixed-Delta} retains the overall search framework of \textbf{BBExplorer} but uses a fixed step size instead of step-size shrinkage.
Together, these baselines allow us to separate the effect of structured search from the additional contribution of progressive refinement.

\paragraph{Scope of comparison.}
Two broader families of methods could in principle be applied to boundary-near input discovery: model-based optimization (e.g., Bayesian optimization) and population-based search (e.g., evolutionary algorithms).
We exclude them for separate reasons.
Model-based methods construct a surrogate predictor from observed evaluations and use it to guide subsequent queries.
In the present setting, environment-side factors and stochastic match conditions are not explicit search variables, and predictor construction under a short release cycle budget would require additional representation and modeling choices, effectively reframing the task as surrogate-based optimization rather than direct boundary testing \cite{ShahriariEtAl2016BO, NIPS2012_05311655}.
Population-based methods maintain a diverse pool of candidates and are effective at covering extreme or disconnected regions of the design space.
However, the scenario targeted here, boundary-onset localization under small, incremental parameter updates, calls for concentrating the evaluation budget near a known balanced region rather than distributing it across distant candidates \cite{deLacerdaEtAl2021ParamControlSLR, LiZhanZhang2022ECSurveyEOP}.
By restricting the comparison to \textit{Random} and \textit{Fixed-Delta}, we isolate two controlled factors, the value of structured search and the additional benefit of shrinkage, without confounding them with modeling or population management decisions \cite{SorokinEtAl2024CID}.
Our focus is not on outperforming general-purpose optimizers, but on isolating the effectiveness of boundary-oriented search under practical testing constraints.


\subsection{Results}
\label{sec:exp-results}
\label{sec:results-turnbased}
\label{sec:exp-lowdim}
\label{sec:exp-realgame}
\label{sec:robustness}
\label{sec:exp-robustness}

We next report the results for low-dimensional performance, higher-dimensional performance, robustness under seed and threshold shifts, and computational cost.

\paragraph{Low-dimensional results.}
In \textit{Turnbased}, we evaluate \textbf{BBExplorer} in 2D and 3D settings, each based on 30 independent trials. We first discuss the 2D case, where the boundary shape is relatively simple. 

In the 2D experiment (ATK $\times$ SPD, 30 trials), \textbf{BBExplorer} achieves a discovery rate of $R = 1.000$, \textit{Fixed-Delta} $R = 0.967$, and \textit{Random} $R = 0.200$.
\textbf{BBExplorer} significantly outperforms \textit{Random} ($p < 0.0001$, $\hat{A}_{12} = 0.900$), confirming the effectiveness of structured search.
Compared with \textit{Fixed-Delta}, the improvement is marginal, indicating that the additional benefit of shrinkage is limited in 2D.

In the 3D experiment, where DEF is added to the search space alongside ATK and SPD, \Cref{tab:turnbased-3d-results} summarizes the results.
\textbf{BBExplorer} maintains $R = 1.000 \pm 0.000$, while \textit{Random} degrades to $R = 0.067 \pm 0.254$, indicating a clear loss of efficiency for unstructured sampling as dimensionality increases.
\textbf{BBExplorer} continues to significantly outperform \textit{Random} ($p < 0.0001$, $\hat{A}_{12} = 0.967$) and also achieves better mean boundary distance ($\bar{D}$) and boundary distance AUC ($A$).
\textit{Fixed-Delta} remains competitive with $R = 0.967 \pm 0.183$, suggesting that the additional benefit of shrinkage remains limited in this setting.

\newcommand{\pmtight}{\mskip1mu\mathord{\pm}\mskip1mu}

\begin{table}[tb]
\centering
\small
\caption{\textit{Turnbased} 3D results (ATK $\times$ SPD $\times$ DEF, 30 trials). $R$: discovery rate; $E$: discovery efficiency; $\bar{D}$: mean boundary distance; $A$: boundary distance AUC.}
\label{tab:turnbased-3d-results}
\setlength{\tabcolsep}{3.5pt}
\begin{tabular}{@{}lcccc@{}}
\toprule
Method & $R$ & $E$ & $\bar{D}$ & $A$ \\
\midrule
\textbf{BBExplorer} & $\textbf{1.000}\pmtight0.000$ & $\textbf{0.76}\pmtight0.85$ & $\textbf{0.0043}\pmtight0.0027$ & $\textbf{0.0830}\pmtight0.0810$ \\
\textit{Fixed-Delta} & $0.967\pmtight0.183$ & $0.69\pmtight0.77$ & $0.0059\pmtight0.0033$ & $0.1250\pmtight0.1171$ \\
\textit{Random} & $0.067\pmtight0.254$ & $0.03\pmtight0.11$ & $0.0766\pmtight0.0783$ & $0.3315\pmtight0.0790$ \\
\bottomrule
\end{tabular}
\end{table}

In summary, \textbf{BBExplorer} reliably discovers boundary-near inputs in low-dimensional settings, with most of the advantage coming from structured search, while shrinkage provides only limited additional benefit.


\paragraph{Higher-dimensional results.}
For \textit{Generals}, results for the higher-dimensional settings (3D, 4D, and 5D) are summarized in \Cref{tab:generals-results}.
We first evaluate a 3D setting by selecting three dimensions from the parameter space of \textit{Generals} and comparing methods over 20 independent trials.
As shown in \Cref{tab:generals-results}, \textbf{BBExplorer} achieves $R = 0.550$ in 3D, confirming that structured search remains effective in this environment.
We next evaluate 4D (20 trials) and 5D (30 trials) settings.
As dimensionality increases, achieving boundary proximity becomes harder.
As shown in \Cref{tab:generals-results}, \textbf{BBExplorer} reaches $R = 0.600$ in 4D and $R = 0.500$ in 5D.
\textbf{BBExplorer} maintains its discovery performance despite the increased difficulty.
By contrast, \textit{Random} degrades to $R = 0.017$ in 5D, showing a near-complete loss of efficiency for unstructured sampling.
\textit{Fixed-Delta} also degrades more strongly than \textbf{BBExplorer} in higher dimensions.
Boundary proximity further supports this pattern: across all dimensions, \textbf{BBExplorer} achieves consistently lower mean boundary distance ($\bar{D}$) than \textit{Fixed-Delta}, indicating tighter convergence to the boundary.
\begin{table}[t]
\centering
\normalsize
\caption{\textit{Generals} results across dimensions. $R$: discovery rate; $\bar{D}$: mean boundary distance (lower is better). Trials: 20 for 3D/4D, 30 for 5D.}
\label{tab:generals-results}
\begin{tabular}{@{}lcccccc@{}}
\toprule
 & \multicolumn{2}{c}{3D} & \multicolumn{2}{c}{4D} & \multicolumn{2}{c}{5D} \\
\cmidrule(lr){2-3} \cmidrule(lr){4-5} \cmidrule(lr){6-7}
Method & $R$ & $\bar{D}$ & $R$ & $\bar{D}$ & $R$ & $\bar{D}$ \\
\midrule
\textbf{BBExplorer} & \textbf{0.550} & \textbf{0.033} & \textbf{0.600} & \textbf{0.049} & \textbf{0.500} & \textbf{0.078} \\
\textit{Fixed-Delta} & 0.000 & 0.120 & 0.200 & 0.092 & 0.067 & 0.108 \\
\textit{Random} & 0.006 & 0.232 & 0.017 & 0.226 & 0.017 & 0.223 \\
\bottomrule
\end{tabular}
\end{table}
The 5D condition serves as the main high-dimensional stress test in this paper.
Although precise boundary localization becomes more difficult there, \textbf{BBExplorer} still preserves discovery performance and behaves more stably than \textit{Fixed-Delta}.
This pattern suggests that, in higher-dimensional settings, shrinkage becomes increasingly important alongside structured search itself.
Overall, \textbf{BBExplorer} continues to outperform \textit{Random}, indicating that structured search remains effective in higher dimensions, while its advantage over \textit{Fixed-Delta} becomes more pronounced, suggesting that step-size shrinkage provides additional benefit as dimensionality increases.

\paragraph{Robustness under seed and threshold shifts.}
\Cref{tab:generalization-test,tab:threshold-diversity} summarize the robustness results.
We consider two robustness analyses.
In the seed-shift analysis, as defined in \Cref{sec:param-winrate}, the same discovered input $P^{(i)}$ in the controllable space is re-evaluated under different random seeds, while all other components of the fixed evaluation condition $\bar{P}$ are held fixed.
In the threshold-shift analysis, $\bar{P}$ is kept fixed, and only the target boundary value $b$ used to define boundary-near inputs in \Cref{sec:testing-goal} is varied.
Together, these analyses address RQ3.
\Cref{tab:generalization-test} summarizes the \textbf{Seed shift}
 results. For the 15 paths that converged in the 5D experiment (out of 30 trials at $R = 0.500$), we reran 1,000 matches for each configuration under 5 different random seeds. The difference between the original search-time estimate and the re-evaluated win rate had a mean of $+0.014$ and a standard deviation of $\pm 0.027$.
The mean inter-seed standard deviation was 0.012, 10 of the 15 paths showed positive diff values, and the worst degradation was only $-0.026$.
These results indicate that the boundary-near inputs discovered by \textbf{BBExplorer} are not overly dependent on a particular seed and remain close to the boundary under shifted stochastic conditions.
\begin{table}[t]
\centering
\normalsize
\caption{Generalization test summary (5D, 15 paths $\times$ 5 seeds $\times$ 1000 games).}
\label{tab:generalization-test}
\begin{tabular}{@{}lccc@{}}
\toprule
Mean diff & Diff SD & Mean inter-seed std & Worst degradation \\
\midrule
$+0.014$ & $\pm 0.027$ & 0.012 & $-0.026$ \\
\bottomrule
\end{tabular}
\end{table}
\Cref{tab:threshold-diversity} summarizes the \textbf{Threshold shift} results. In \Cref{sec:testing-goal}, boundary-near inputs are defined relative to a target boundary value $b \in \{l,h\}$. 
Here we compare the additional settings $b=0.40$ and $b=0.50$,
each evaluated over 10 trials, with the main $b=0.60$ setting,
evaluated over 30 trials, to assess stability under different boundary definitions.
\textbf{BBExplorer} achieved a mean discovery rate of $R = 0.533$ across the three thresholds, whereas \textit{Fixed-Delta} and \textit{Random} showed considerably lower and less consistent results.
The mean boundary distance of \textbf{BBExplorer} decreased from $b = 0.60$ to $b = 0.40$, indicating that localization became easier at lower thresholds.
These results confirm that \textbf{BBExplorer} remains stable across boundary definitions and that shrinkage continues to provide an additional benefit under threshold shifts.
\begin{table}[t]
\centering
\normalsize
\caption{Target boundary value variation in 5D. The additional settings
$b=0.40$ and $b=0.50$ use 10 trials each, while the main $b=0.60$
setting uses 30 trials. $R$: discovery rate; $\bar{D}$: mean boundary
distance (lower is better).}
\label{tab:threshold-diversity}
\begin{tabular}{@{}lcccccc@{}}
\toprule
 & \multicolumn{2}{c}{$b=0.40$} & \multicolumn{2}{c}{$b=0.50$} & \multicolumn{2}{c}{$b=0.60$} \\
\cmidrule(lr){2-3} \cmidrule(lr){4-5} \cmidrule(lr){6-7}
Method & $R$ & $\bar{D}$ & $R$ & $\bar{D}$ & $R$ & $\bar{D}$ \\
\midrule
\textbf{BBExplorer} & \textbf{0.600} & \textbf{0.013} & \textbf{0.500} & \textbf{0.025} & \textbf{0.500} & \textbf{0.078} \\
\textit{Fixed-Delta} & 0.200 & 0.062 & 0.300 & 0.100 & 0.067 & 0.108 \\
\textit{Random} & 0.043 & 0.179 & 0.029 & 0.189 & 0.017 & 0.223 \\
\bottomrule
\end{tabular}
\end{table}

\paragraph{Computational cost.}
From a practical standpoint, \textbf{BBExplorer} also remained computationally efficient in the high-dimensional setting.
In the main 5D experiment, the average number of matches per path was approximately 9,000 for \textbf{BBExplorer} and approximately 15,000 for \textit{Fixed-Delta}.
The total execution time for all additional experiments\footnote{All experiments were conducted on a machine equipped with an AMD Ryzen 9 7900 CPU, 64GB RAM, and an NVIDIA GeForce RTX 4070 Ti SUPER GPU.} was approximately 2.7 hours: about 65 minutes for the 5D 30-trial experiment, about 58 minutes for the 3D and 4D experiments, and about 40 minutes for the threshold-shift experiments.
These figures confirm that \textbf{BBExplorer} operates within a practical computational budget while maintaining its discovery advantage over \textit{Fixed-Delta}.

\section{Discussion}
\label{sec:discussion}

\subsection{Answers to Research Questions}
\label{sec:rq-answers}


We answer the three research questions based on the results in \Cref{sec:experiments}.

\paragraph{RQ1 (Effectiveness). Strongly supported.}
In the \textit{Turnbased} 3D setting, \textbf{BBExplorer} achieves $R = 1.000$ compared with $R = 0.067$ for \textit{Random} ($p < 0.0001$, $\hat{A}_{12} = 0.967$), along with a substantially smaller mean boundary distance ($\bar{D} = 0.0043$ vs.\ $0.0766$).
In the \textit{Generals} 5D setting, \textbf{BBExplorer} maintains $R = 0.500$, whereas \textit{Random} degrades to $R = 0.017$ (\Cref{tab:turnbased-3d-results,tab:generals-results}).
These results confirm that structured search significantly improves boundary discovery performance.

\paragraph{RQ2 (Scalability). Largely supported.}
In the \textit{Generals} environment, \textbf{BBExplorer} achieves $R = 0.550$ in 3D, $R = 0.600$ in 4D, and $R = 0.500$ in 5D, while \textit{Fixed-Delta} remains at $R \leq 0.200$ and \textit{Random} at $R \leq 0.017$ across all dimensions (\Cref{tab:generals-results}).
These results indicate that \textbf{BBExplorer} retains meaningful discovery performance as dimensionality increases, although precise boundary localization becomes more challenging in 5D.

\paragraph{RQ3 (Robustness). Supported.}
Re-evaluation of the 15 converged 5D paths under five different random seeds shows a mean win-rate difference of $+0.014$ ($\pm 0.027$) from the original search-time estimates, with a worst-case degradation of only $-0.026$ (\Cref{tab:generalization-test}).
Under threshold shifts, \textbf{BBExplorer} achieves a mean discovery rate of $R = 0.533$ across three boundary values ($b = 0.40, 0.50, 0.60$), while \textit{Fixed-Delta} and \textit{Random} remain substantially lower (\Cref{tab:threshold-diversity}).
These results indicate that the discovered boundary-near inputs remain stable under both stochastic variation and changes in boundary definitions.

\paragraph{Summary.}
Overall, the results support the effectiveness of structured search for boundary discovery, show that step-size shrinkage provides additional benefits as dimensionality increases, and demonstrate that the discovered boundary-near inputs remain robust under stochastic variation and changes in boundary definitions.

\subsection{Why BBExplorer Works}
\label{sec:why-it-works}

The experimental results show that \textbf{BBExplorer} consistently outperforms the baselines, achieving strong performance even in the most demanding \textit{Generals} 5D setting and maintaining stability under both seed and threshold shifts (\Cref{tab:generals-results,tab:generalization-test,tab:threshold-diversity}).
We attribute these outcomes to three design elements, each addressing a distinct challenge.

\paragraph{Multi-directional candidate generation.}
At each iteration, \textbf{BBExplorer} probes both directions of every axis simultaneously.
This makes the search responsive to local boundary shape, regardless of which axis is most relevant at a given point.
As dimensionality increases, boundary orientation becomes harder to predict, and searching along a single axis or at random is more likely to miss the locally informative direction.
Multi-directional probing reduces this risk, contributing to the sustained discovery performance observed across dimensions.

\paragraph{Two-stage screening.}
The screening stage filters candidates using a small number of matches before committing the full evaluation budget.
This design allocates computation to the most promising candidates rather than distributing it evenly.
It also provides robustness against stochastic match outcomes: a candidate that passes screening is less likely to be a noise artifact.
This is consistent with the seed-shift analysis in \Cref{sec:exp-robustness}, where re-evaluation under different random seeds produced only minor deviations from the original estimates.

\paragraph{Step-size shrinkage.}
A fixed step size must compromise between rapid progression toward the boundary and precise localization.
Shrinkage removes this trade-off by allowing large steps when the search is far from the boundary and small steps when it is close.
This transition is particularly important in higher dimensions, where the distance to the boundary can vary more across axes.
The contrast with \textit{Fixed-Delta}, which shares the same search framework but lacks shrinkage, directly isolates this contribution.
These three elements address different aspects of the problem, namely multi-directional exploration, budget allocation under non-determinism, and progressively finer boundary localization.
Their combination enables the consistent performance observed across the experimental conditions.

\subsection{Limitations}
\label{sec:failure-cases}

In this subsection, we examine the main limitations of \textbf{BBExplorer}.
Whereas \Cref{sec:why-it-works} focused on why \textbf{BBExplorer} works, this subsection centers on the main limitations observed in the demanding \textit{Generals} setting.

\paragraph{Non-convergence.}
When starting points are far from the boundary, or when screening-stage variance is high, promising directions may fail to advance to full evaluation.
As a result, the search may not reach the boundary-near region within the available iterations.
This non-convergence becomes more visible in higher-dimensional settings.

\paragraph{Fixed simulation counts.}
The current implementation uses fixed simulation counts
$N_{\mathrm{scr}}$ and $N_{\mathrm{full}}$ for screening and full
evaluation.
A natural extension is a confidence-interval-driven stopping rule that
adapts these counts based on the estimated distance from the target
boundary, using a looser interval during screening and a tighter
interval during full evaluation.
We leave this dynamic allocation for future work.

\paragraph{Step-size overshooting.}
When a starting point already lies near the boundary, a large initial
step size may jump over the boundary-near region in early iterations.
Highly irregular boundary shapes can likewise cause the search to leave
the useful neighborhood before the step size has sufficiently decayed.

\paragraph{Axis-aligned probing.}
Strong parameter interactions may require joint movements across multiple axes that axis-aligned probing cannot represent directly.
Although \textbf{BBExplorer} evaluates both directions of every axis at each iteration, it can still miss boundary structures whose locally informative direction is not aligned with any single axis.
This limitation reflects the greedy, axis-aligned nature of the search and suggests extending candidate generation beyond axis-aligned moves.
Comparing \textbf{BBExplorer} with non-greedy or population-based search strategies, therefore, remains an important direction for future work.

\paragraph{Single-path coverage.}
\textbf{BBExplorer} follows a single convergent path from each starting point.
Accordingly, the discovered boundary-near inputs are limited to the portion of the boundary reachable along those paths.
This limitation should be interpreted as a coverage constraint rather than outright failure.
The goal of the approach is not exhaustive boundary mapping, but practical acquisition of useful boundary-near inputs under a finite simulation budget.

\paragraph{Incremental scope.}
This work primarily targets scenarios where incremental parameter updates to an existing design trigger balance disruption.
It is less suited to extreme parameter configurations that lie far from the known balanced region, including highly skewed settings that occupy disconnected regions of the parameter space.
Such cases may require search designs beyond continuous step-size shrinkage alone.
At the same time, increasing the diversity of the starting point set offers a practical way to extend coverage toward such regions.
Despite these limitations, \textbf{BBExplorer} still demonstrates practical effectiveness in discovering useful boundary-near inputs across the \textit{Turnbased} (2D/3D) and \textit{Generals} (3D--5D) settings evaluated in \Cref{sec:experiments}.
The cases discussed above also point to clear directions for extension, including richer candidate generation and broader multi-path coverage.

\subsection{Threats to Validity}
\label{sec:threats}

\paragraph{Internal validity.}
The results depend on the evaluation protocol used in \Cref{sec:experiments}.
Hyperparameters such as the screening and full evaluation counts, the initial step size, and the decay factor can influence the search behavior.
Moreover, win-rate estimation involves Monte Carlo variance, and this residual variance cannot be completely eliminated near the boundary.
Although all methods were compared under the same conditions and we additionally performed seed-shift analysis, the quantitative results may still vary under other configurations.
That said, the consistent ranking across both environments and the stability confirmed by the seed-shift analysis suggest that the main conclusions are not merely an artifact of the specific experimental setup used here.

\paragraph{External validity.}
The experiments are limited to two environments, \textit{Turnbased} and \textit{Generals}, and to settings of up to five dimensions.
These two environments nonetheless cover a range of complexity,
from a controlled synthetic setting to an environment based on
a real-world online strategy game.
Additional studies are needed before generalizing the conclusions to higher-dimensional spaces, mixed-type parameter spaces, or commercial live-ops games.
Within this scope, BBExplorer is not tied to a specific game genre.
As formulated in \Cref{sec:problem}, it applies when developers can define
controllable parameters $P$, a representative evaluation condition
$\bar{P}$, and a measurable balance objective such as win rate.
For real-time, human-input-heavy, or many-parameter games, the main
challenge is defining representative $\bar{P}$ conditions and selecting
or grouping controllable parameters, rather than changing the
boundary-discovery formulation itself.
The comparison set is limited to unstructured sampling and a core ablation.
This design isolates structured search and step-size shrinkage, but it does
not establish superiority over population-based, quality-diversity, or
surrogate-based optimizers.
A fair comparison would require adapting those methods to near-boundary
discovery under the same stochastic match budget.
Under that setting, BBExplorer concentrates evaluations along individual
search paths, whereas population-based methods may cover a broader set of
boundary-near regions.
Such comparisons remain important future work.

In addition, $\bar{P}$ is treated as a fixed evaluation condition throughout the experiments.
Although the seed-shift analysis in \Cref{sec:exp-robustness} re-evaluated convergence under different random outcomes, opp\-onent-side strategy diversity was not systematically varied.
Evaluation under broader opponent variation and strategy diversity remains an important direction for further validation.
Separately, the seed-shift results provide initial evidence that the discovered boundary-near inputs are not overly dependent on a single stochastic condition.

\newcommand{\approachname}{\textbf{BBExplorer}}
\section{Related Work}


\subsection{Game Balance Testing}
Various parameter optimization approaches that leverage reinforcement learning (RL) and genetic algorithms have been proposed in game balance studies. For instance, approaches such as LUDUS utilize evolutionary computation to optimize attribute values in card games, while GEEvo focuses on optimizing resource flows within game economies \cite{Budijono2022LUDUS,Rupp2024GEEvo}. Additionally, RL-based approaches for evaluating and optimizing map configurations and deep learning-based behavioral modeling to assess balance have been explored \cite{Rupp2025SimDriven,Pfau2020}. These conventional approaches primarily aim to identify a single set of optimal parameter values that satisfy specific evaluation metrics defined by developers. However, they typically require a high number of iterations to converge and do not focus on identifying the specific boundaries at which game balance begins to break down.

In contrast, \approachname{} aims to efficiently identify the boundaries at which game balance becomes disrupted, along with their surrounding regions. 
In Games-as-a-Service settings, where continuous post-launch updates are required \cite{Dubois2021}, developers must repeatedly adjust parameters to prevent imbalance. 
Understanding these boundaries is therefore more critical than identifying a single optimal configuration.

\subsection{Search-Based Software Testing}


From a search-based software testing (SBST) perspective, game balance adjustment can be framed as a search problem over parameter configurations \cite{Bertolino2007, McMinn2004SBSTSurvey}. 
Typical SBST techniques focus on efficiently exploring the input space to detect failures with a minimal number of test executions or to achieve high coverage, thereby improving software quality.

In contrast, \approachname{} focuses on boundary-oriented exploration under non-deterministic evaluation. 
Rather than emphasizing failure detection or coverage, it targets parameter configurations near the boundary where acceptable behavior transitions into imbalance. 
This shift reflects the need to characterize regions of behavioral change rather than locate isolated faults or maximize coverage.

\subsection{Automated Game Testing}

In game development testing, instead of performing low-level tests such as unit tests, higher-level tests such as playtesting, in which testers play the game, are given priority \cite{murphy2014cowboys, Politowski2021}. Automated approaches to playtesting have been actively studied, though the gap between academic solutions and the actual needs of game developers remains significant \cite{Politowski2022}. 

Fuzzing is an automated testing approach in which an initial input is repeatedly mutated to generate new inputs, termed fuzz, that are likely to cause issues \cite{manes2018art}. While it is generally used in the security field, in recent years it has been adopted in various fields, such as autonomous driving systems \cite{Guo2024} and physics simulation engines \cite{Xiao2023}. AFL\footnote{\url{https://lcamtuf.coredump.cx/afl/}} is a representative coverage-guided grey-box fuzzer with a proven track record of detecting issues in various software. It measures code coverage for each input and uses this information to guide the generation of subsequent fuzz. Aschermann et al. proposed IJON to enhance AFL and applied it to Super Mario Bros to automate playtesting \cite{aschermann2020IJON}. They demonstrated that conventional fuzzing based on code coverage performs inadequately for software with complex state transitions, such as video games. Furthermore, they proposed an extended tool named IJON to improve AFL based on annotations. Consequently, they demonstrated the alleviation of state explosion. Also, BiFuzz is a two-stage fuzzer that automates playtesting for open-world games developed using game engines \cite{Kato2024}. Both tools focus on playtesting automation using fuzzing. However, IJON aims to transition the game to a specific state, while BiFuzz aims to detect a character stuck issue as a common problem in open-world games.

Agent-based and learning-based approaches offer alternatives to fuzzing. Fan et al. proposed a method using machine learning for automated playing in open-world games \cite{fan2022minedojo,wang2023voyager}. This research aims to perform a variety of tasks within the game. Similarly, Chen et al. proposed MIMIC, which uses a large language model to simulate diverse player personality traits for automated game testing \cite{chen2025mimic}. Prasetya et al. developed aplib for programming intelligent test agents \cite{Prasetya2020} and later extended it to smart playtesting with dynamic goal structures \cite{Shirzadehhajimahmood2021, Prasetya2026}. Stahlke et al. proposed PathOS, which simulates player navigation to support level design evaluation \cite{Stahlke2020}. Prasetya et al. also demonstrated automated navigation and exploration in 3D game environments \cite{Prasetya2020Nav}.

Other approaches operate at the pixel level or use RL. Zheng et al. proposed Wuji, which combines evolutionary algorithms with deep reinforcement learning for automatic online combat game testing \cite{Wuji2019}. Liu et al. proposed Inspector, a pixel-based tool that integrates exploration, detection, and investigation for automated game testing \cite{Inspector2022}. From a game software engineering perspective, Casamayor et al. addressed bug localization in video games using evolving simulations \cite{Casamayor2022}.

Beyond individual testing techniques, Kato et al. use open-world games as
an extreme case to examine software testing under uncertainty.
They identify inexhaustible behavior spaces, non-deterministic execution
outcomes, elusive behavioral boundaries, and unstable test oracles as
recurring challenges, and argue for selective exploration of behaviorally
meaningful regions under cost constraints and explicit treatment of
execution variability \cite{FSE2026Kato}.

Within this broader perspective, \approachname{} focuses on a concrete
objective: boundary-oriented game balance regression testing.
Whereas prior automated game testing methods primarily focus on reaching
specific game states, detecting anomalies, or executing gameplay scenarios,
\approachname{} searches for parameter configurations near the boundary
between balanced and unbalanced gameplay under a finite simulation budget.
It therefore characterizes regions of behavioral transition through repeated
stochastic evaluation rather than targeting predefined states or discrete
issues.

\section{Conclusion}
\label{sec:conclusion}

This paper formulated balance regression testing in competitive games as a boundary-discovery problem under stochastic evaluation and limited simulation budgets.
We introduced \textbf{BBExplorer}, which combines multi-directional candidate generation, two-stage screening, and adaptive step-size shrinkage.

Our experiments show that \textbf{BBExplorer} consistently identifies boundary-near inputs more effectively than baseline methods, and maintains this advantage in higher-dimensional settings and under seed and threshold shifts.
These results suggest that structured search combined with adaptive refinement is effective for boundary localization under non-deterministic conditions.

Future work includes extending the search strategy and evaluating the approach in more complex parameter spaces and real-world game settings.


\section*{Data Availability Statement}
The replication package, including source code and experimental artifacts, is available at \url{https://doi.org/10.5281/zenodo.21076109}.





\begin{acks}
We thank Dr. Masanari Kondo and Mr. Sota Nakashima at Kyushu University for their valuable comments.
This work was supported by JSPS KAKENHI Grant Numbers JP24K02923, JP23K28065.
The authors utilized generative AI tools, including ChatGPT and Claude, to assist with the English translation, text refinement, literature organization, and LaTeX formatting of this paper. All generated content was thoroughly reviewed and edited by the authors, who take full responsibility for the final manuscript.
\end{acks}

\bibliographystyle{ACM-Reference-Format}
\bibliography{sample-base}

\end{CJK*}
\end{document}